\documentclass[final]{IEEEtran}
\IEEEoverridecommandlockouts
\usepackage{cite}
\usepackage{amsmath,amssymb,amsfonts}
\usepackage{algorithmic}
\usepackage{graphicx}
\usepackage{textcomp}
\usepackage{xcolor}
\usepackage{array}
\usepackage{hyperref}
\usepackage{etoolbox}
\appto\UrlBreaks{\do\-}

\def\BibTeX{{\rm B\kern-.05em{\sc i\kern-.025em b}\kern-.08em
    T\kern-.1667em\lower.7ex\hbox{E}\kern-.125emX}}
\begin{document}

\title{AI Accelerator Survey and Trends\\
\thanks{This material is based upon work supported by the Assistant Secretary of Defense for Research and Engineering under Air Force Contract No. FA8702-15-D-0001. Any opinions, findings, conclusions or recommendations expressed in this material are those of the author(s) and do not necessarily reflect the views of the Assistant Secretary of Defense for Research and Engineering.}
}

\author{\IEEEauthorblockN{Albert Reuther, Peter Michaleas, Michael Jones, Vijay Gadepally, Siddharth Samsi, and Jeremy Kepner} \\
\IEEEauthorblockA{\textit{MIT Lincoln Laboratory Supercomputing Center} \\
Lexington, MA, USA \\
\{reuther,pmichaleas,michael.jones,vijayg,sid,kepner\}@ll.mit.edu}
}

\maketitle

\begin{abstract}

%%% Needs editing
Over the past several years, new machine learning accelerators were being announced and released every month for a variety of applications from speech recognition, video object detection, assisted driving, and many data center applications. This paper updates the survey of AI accelerators and processors from past two years. This paper collects and summarizes the current commercial accelerators that have been publicly announced with peak performance and power consumption numbers. The performance and power values are plotted on a scatter graph, and a number of dimensions and observations from the trends on this plot are again discussed and analyzed. This year, we also compile a list of benchmarking performance results and compute the computational efficiency with respect to peak performance. 

\end{abstract}


\section{Introduction}

Over the past several years, startups and established technology companies have been announcing, releasing, and deploying a wide variety of artificial intelligence (AI) and machine learning (ML) accelerators. The focus of these accelerators has been on accelerating deep neural network (DNN) models, and the application space spans from very low power embedded voice recognition to data center scale training. The announcements of new accelerators has slowed in the past year, but the competition for defining markets and application areas continues. 
This drive to developing and deploying accelerators has been part of a much larger industrial and technology shift in modern computing.

\begin{figure}[th]
    \centering
    \includegraphics[width=3in]{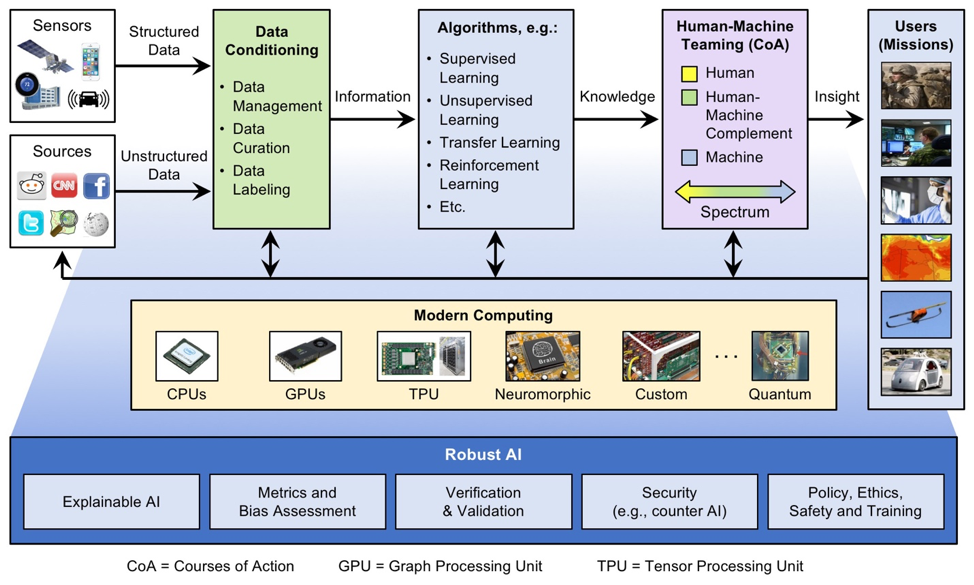}
    \caption{Canonical AI architecture consists of sensors, data conditioning, algorithms, modern computing, robust AI, human-machine teaming, and users (missions). Each step is critical in developing end-to-end AI applications and systems.}
    \label{fig:architecture}
  \end{figure}

AI ecosystems bring together a components from embedded computing (edge computing), traditional high performance computing (HPC), and high performance data analysis (HPDA) that must work together to effectively provide capabilities for use by decision makers, warfighters, and analysts~\cite{gadepally2019enabling}. 
Figure~\ref{fig:architecture} captures an architectural overview of such end-to-end AI solutions and their components. 
On the left side of Figure~\ref{fig:architecture}, structured and unstructured data sources provide different views of entities and/or phenomenology. 
These raw data products are fed into a data conditioning step in which they are fused, aggregated, structured, accumulated, and converted into information. The information generated by the data conditioning step feeds into a host of supervised and unsupervised algorithms such as neural networks, which extract patterns, predict new events, fill in missing data, or look for similarities across datasets, thereby converting the input information to actionable knowledge. This actionable knowledge is then passed to human beings for decision-making processes in the human-machine teaming phase. The phase of human-machine teaming provides the users with useful and relevant insight turning knowledge into actionable intelligence or insight. 

%%% Ended here

Underpinning this system are modern computing systems. Moore's law trends have ended~\cite{theis2017end}, as have a number of related laws and trends including Denard's scaling (power density), clock frequency, core counts, instructions per clock cycle, and instructions per Joule (Koomey's law)~\cite{horowitz2014computing}. Taking a page from the system-on-chip (SoC) trends first seen in automotive and smartphones, advancements and innovations are still progressing by developing and integrating accelerators for often-used operational kernels, methods, or functions.  These accelerators are designed with a different balance between performance and functional flexibility. This includes an explosion of innovation in deep machine learning processors and accelerators~\cite{leiserson2020theres,thompson2021decline,hennessy2019new,dally2020domain,lecun2019deep}. Understanding the relative benefits of these technologies is of particular importance to applying AI to domains under significant constraints such as size, weight, and power, both in embedded applications and in data centers.

% Add material about domain specific architectures \cite{dally2020domain} \cite{jouppi2020domain}. Include articles about ASIC Clouds, cryptocurrency mining, video transcoding

% General purpose processors ? CPUs, Parallel engines (vector processors) ? GPUs, Domain specific architectures (still programmable), algorithm/protocol specific chips/chiplets/circuit IP (dataflow, not programmable) -- the circuit is the program

This paper is an update to IEEE-HPEC papers from the past two years~\cite{reuther2020survey,reuther2019survey}. 
As in past years, we will review a few topics pertinent to understanding the capabilities of the accelerators. We must discuss the types of neural networks for which these ML accelerators are being designed; the distinction between neural network training and inference; the numerical precision with which the neural networks are being used for training and inference, and how neuromorphic and optical accelerators fit into the mix: 

\begin{itemize}

\item Types of Neural Networks -- While AI and machine learning encompass a wide set of statistics-based technologies~\cite{gadepally2019enabling}, this paper continues with last year's focus on accelerators and processors that are geared toward deep neural networks (DNNs) and convolutional neural networks (CNNs) as they are quite computationally intense~\cite{canziani2016analysis}.  

\item Neural Network Training versus Inference -- As was explained in the previous two survey, the survey focuses on accelerators and processors for inference for a variety of reasons including that defense and national security AI/ML edge applications rely heavily on inference. 

\item Numerical Precision -- We will consider all of the numerical precision types that an accelerator supports, but for most of them, their best inference performance is in int8 or fp16/bf16 (IEEE 16-bit floating point or Google's 16-bit brain float). But as can be seen in Figure~\ref{fig:PeakPerformancePower}, peak performance has been reported for many different numerical formats. 

\item Neuromorphic Computing and Photonic Computing -- For this year's survey, we are going to take a pause on most of the neuromorphic computing and photonic computing accelerators. Several new accelerators have been released, but none of these companies have released peak performance and peak power numbers for them. There have been some relative comparisons of neuromorphic processors to conventional accelerators (e.g., \cite{ward2020intel}), but there have been no hard numbers. Perhaps next year we will start seeing actual performance numbers that we can incorporate in this survey. 

\end{itemize}

There are many surveys~\cite{lindsey1995survey,liao2001neural,misra2010artificial,sze2017efficient,sze2020efficient,langroudi2019digital,chen2020survey,wang2019deep,khan2020ai,rueckert2020digital} and other papers that cover various aspects of AI accelerators; this multi-year survey effort and this paper focus on gathering a comprehensive list of AI accelerators with their computational capability, power efficiency, and ultimately the computational effectiveness of utilizing accelerators in embedded and data center applications. Along with this focus, this paper mainly compares neural network accelerators that are useful for government and industrial sensor and data processing applications. 

\section{Survey of Processors}

%\begin{landscape}
\begin{figure*}[htb]
    \centering
    \includegraphics[width=\textwidth]{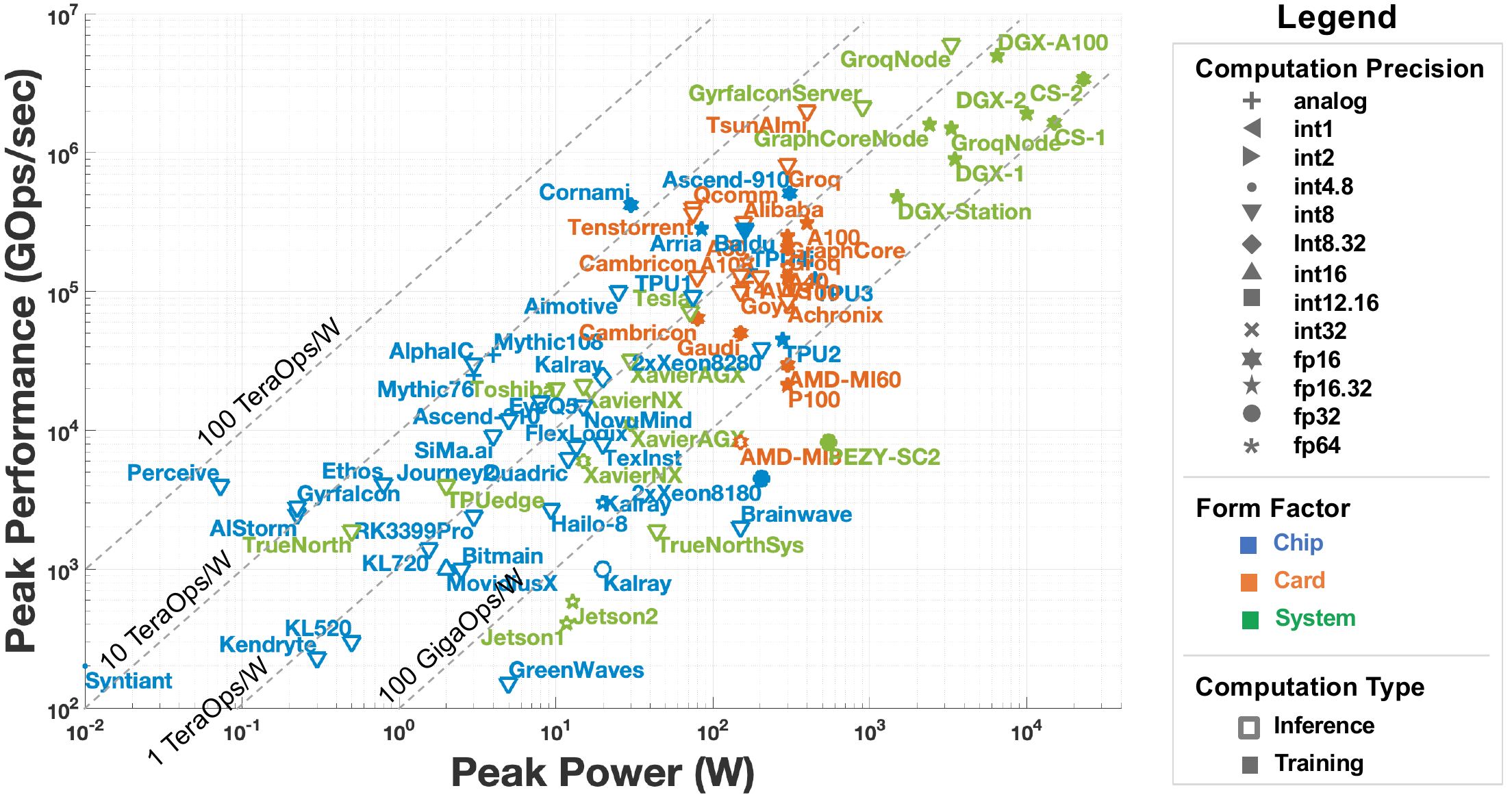}
    \caption{Peak performance vs. power scatter plot of publicly announced AI accelerators and processors.}
    \label{fig:PeakPerformancePower}
  \end{figure*}
%\end{landscape}

Many recent advances in AI can be at least partly credited to advances in computing hardware~\cite{krizhevsky2012imagenet,jouppi2018domain}, enabling computationally heavy machine-learning algorithms and in particular DNNs. This survey gathers performance and power information from publicly available materials including research papers, technical trade press, company benchmarks, etc. While there are ways to access information from companies and startups (including those in their silent period), this information is intentionally left out of this survey; such data will be included in this survey when it becomes publicly available. The key metrics of this public data are plotted in Figure~\ref{fig:PeakPerformancePower}, which graphs recent processor capabilities (as of July 2021) mapping peak performance vs. power consumption. 

The x-axis indicates peak power, and the y-axis indicate peak giga-operations per second (GOps/s), both on a logarithmic scale. Note the legend on the right, which indicates various parameters used to differentiate computing precisions, form factors, and inference/training. The computational precision of the processing capability is depicted by the geometric shape used; the computational precision spans from analog and single-bit int1 to four-byte int32 and two-byte fp16 to eight-byte fp64. The precisions that show two types denotes the precision of the multiplication operations on the left and the precision of the accumulate/addition operations on the right (for example, fp16.32 corresponds to fp16 for multiplication and fp32 for accumulate/add). The form factor is depicted by  color; this is important for showing how much power is consumed, but also how much computation can be packed onto a single chip, a single PCI card, and a full system. Blue corresponds to a single chip; orange corresponds to a card (note that they all are in the 200-400 Watt zone); and green corresponds to entire systems (single node desktop and server systems). This survey is limited to single motherboard, single memory-space systems. Finally, the hollow geometric objects are peak performance for inference-only accelerators, while the solid geometric figures are performance for accelerators that are designed to perform both training and inference. 

The survey begins with the same scatter plot that we have compiled for the past two years~\cite{reuther2020survey,reuther2019survey}. To save space, we have summarized some of the important metadata of the accelerators, cards, and systems in Table~\ref{tab:acceleratorlist}, including the label used in Figure~\ref{fig:PeakPerformancePower} for each of the points on the graph; many of the points were brought forward from last year's plot, and some details of those entries are in~\cite{reuther2020survey}. There are several additions which we will cover below. 
In Table~\ref{tab:acceleratorlist}, most of the columns and entries are self explanatory. However, there are two Technology entries that may not be: dataflow and PIM. Dataflow processors are custom-designed processors for neural network inference and training. Since neural network training and inference computations can be entirely deterministically laid out, they are amenable to dataflow processing in which computations, memory accesses, and inter-ALU communications actions are explicitly programmed or ``placed-and-routed'' onto the computational hardware. Processor in memory (PIM) is an analog computing technology that augments flash memory circuits with in-place analog multiply-add capabilities. Please refer to the references for the Mythic and Gyrfalcon accelerators for more details on this innovative technology. 

\begingroup
\setlength{\tabcolsep}{2pt} % Default value: 6pt

\begin{table*}
  \centering
  \scriptsize
  \caption{List of accelerator labels for plots.}
  \label{tab:acceleratorlist}

\begin{tabular}{| l | l | c | c | c | c |} \hline 

 \textbf{Company} & \textbf{Product} & \textbf{Label} & \textbf{Technology} & \textbf{Form Factor} & \textbf{References} \\ \hline 
     Achronix & VectorPath S7t-VG6 & Achronix & dataflow & Card & \cite{roos2019fpga}  \\ \hline  
     Aimotive & aiWare3 & Aimotive & dataflow & Chip & \cite{aimotive2018aiware3}  \\ \hline  
     AIStorm	 & AIStorm & AIStorm & dataflow & Chip & \cite{merritt2019aistorm}  \\ \hline  
     Alibaba & Alibaba & Alibaba & dataflow & Card & \cite{peng2019alibaba}  \\ \hline  
     AlphaIC & RAP-E & AlphaIC & dataflow & Chip & \cite{clarke2018indo}  \\ \hline  
     Amazon & Inferentia & AWS & dataflow & Card & \cite{hamilton2018aws,cloud2020deep}  \\ \hline  
     AMD & Radeon Instinct MI6 & AMD-MI8 & GPU & Card & \cite{exxactcorp2017taking}  \\ \hline  
     AMD & Radeon Instinct MI60 & AMD-MI60 & GPU & Card & \cite{smith2018amd}  \\ \hline  
     ARM & Ethos N77 & Ethos & dataflow & Chip & \cite{schor2020arm}  \\ \hline  
     Baidu & Baidu Kunlun 818-300 & Baidu & dataflow & Chip & \cite{ouyang2021kunlun,merritt2018baidu,duckett2018baidu}  \\ \hline  
     Bitmain & BM1880 & Bitmain & dataflow & Chip & \cite{wheeler2019bitmain}  \\ \hline  
     Blaize & El Cano & Blaize & dataflow & Card & \cite{demler2020blaize}  \\ \hline  
     Cambricon & MLU100 & Cambricon & dataflow & Card & \cite{wu2018chinese,cutress2018cambricon}  \\ \hline  
     Canaan & Kendrite K210 & Kendryte & CPU & Chip & \cite{gwennap2019kendryte}  \\ \hline  
     Cerebras & CS-1 & CS-1 & dataflow & System & \cite{hock2019introducing}  \\ \hline  
     Cerebras & CS-2 & CS-2 & dataflow & System & \cite{trader2021cerebras}  \\ \hline  
     Cornami & Cornami & Cornami & dataflow & Chip & \cite{cornami2020cornami}  \\ \hline  
     Enflame & Cloudblazer T10 & Enflame & CPU & Card & \cite{clarke2019globalfoundries}  \\ \hline  
     Flex Logix & InferX X1 & FlexLogix & dataflow & Chip & \cite{mehta2020performance}  \\ \hline  
     Google & TPU Edge & TPUedge & dataflow & System & \cite{tpu2019edge}  \\ \hline  
     Google & TPU1 & TPU1 & dataflow & Chip & \cite{jouppi2020domain,teich2018tearing}  \\ \hline  
     Google & TPU2 & TPU2 & dataflow & Chip & \cite{jouppi2020domain,teich2018tearing}  \\ \hline  
     Google & TPU3 & TPU3 & dataflow & Chip & \cite{jouppi2021ten,jouppi2020domain,teich2018tearing}  \\ \hline  
     Google & TPU4i & TPU4i & dataflow & Chip & \cite{jouppi2021ten}  \\ \hline  
     GraphCore & C2 & GraphCore & dataflow & Card & \cite{gwennap2020groq,lacey2017preliminary}  \\ \hline  
     GraphCore & C2 & GraphCoreNode & dataflow & System & \cite{graphcore2020dell}  \\ \hline  
     GreenWaves & GAP9 & GreenWaves & dataflow & Chip & \cite{greenwaves2020gap,turley2020gap9}  \\ \hline  
     Groq & Groq Node & GroqNode & dataflow & System & \cite{hemsoth2020groq}  \\ \hline  
     Groq & Tensor Streaming Processor & Groq & dataflow & Card & \cite{gwennap2020groq,abts2020think}  \\ \hline  
     Gyrfalcon & Gyrfalcon & Gyrfalcon & PIM & Chip & \cite{ward2019gyrfalcon}  \\ \hline  
     Gyrfalcon & Gyrfalcon & GyrfalconServer & PIM & System & \cite{hpcwire2020solidrun}  \\ \hline  
     Habana & Gaudi & Gaudi & dataflow & Card & \cite{gwennap2019habanagaudi,medina2020habana}  \\ \hline  
     Habana & Goya HL-1000 & Goya & dataflow & Card & \cite{gwennap2019habanagoya,medina2020habana}  \\ \hline  
     Hailo & Hailo & Hailo-8 & dataflow & Chip & \cite{ward2019details}  \\ \hline  
     Horizon Robotics & Journey2 & Journey2 & dataflow & Chip & \cite{horizon2020journey}  \\ \hline  
     Huawei HiSilicon & Ascend 310 & Ascend-310 & dataflow & Chip & \cite{huawei2020ascend310}  \\ \hline  
     Huawei HiSilicon & Ascend 910 & Ascend-910 & dataflow & Chip & \cite{huawei2020ascend910}  \\ \hline  
     IBM & TrueNorth & TrueNorth & neuromorphic & System & \cite{feldman2016ibm,esser2016convolutional,akopyan2015truenorth}  \\ \hline  
     IBM & TrueNorth & TrueNorthSys & neuromorphic & System & \cite{feldman2016ibm,esser2016convolutional,akopyan2015truenorth}  \\ \hline  
     Intel & Arria 10 1150 & Arria & FPGA & Chip & \cite{abdelfattah2018dla,hemsoth2018intel}  \\ \hline  
     Intel & Mobileye EyeQ5 & EyeQ5 & dataflow & Chip & \cite{demler2020blaize}  \\ \hline  
     Intel & Movidius Myriad X & MovidiusX & manycore & Chip & \cite{hruska2017new}  \\ \hline  
     Intel & Xeon Platinum 8180 & 2xXeon8180 & multicore & Chip & \cite{degelas2019intel,cpuworld2020xeon8180}  \\ \hline  
     Intel & Xeon Platinum 8280 & 2xXeon8280 & multicore & Chip & \cite{degelas2019intel,cpuworld2020xeon8280}  \\ \hline  
     Kalray & Coolidge & Kalray & manycore & Chip & \cite{dupont2019kalray, clarke2020nxp}  \\ \hline  
     Kneron & KL520 Neural Processing Unit & KL520 & dataflow & Chip & \cite{ward2020kneron}  \\ \hline  
     Kneron & KL720 & KL720 & dataflow & Chip & \cite{ward2021kneron}  \\ \hline  
     Microsoft & Brainwave & Brainwave & dataflow & Chip & \cite{morgan2017drilling}  \\ \hline  
     Mythic & M1076 & Mythic76 & PIM & Chip & \cite{ward2021mythic,hemsoth2018mythic,fick2018mythic}  \\ \hline  
     Mythic & M1108 & Mythic108 & PIM & Chip & \cite{ward2021mythic,hemsoth2018mythic,fick2018mythic}  \\ \hline  
     NovuMind & NovuTensor & NovuMind & dataflow & Chip & \cite{freund2019novumind,yoshida2018novumind}  \\ \hline  
     NVIDIA & Ampere A100 & A100 & GPU & Card & \cite{krashinsky2020nvidia}  \\ \hline  
     NVIDIA & Ampere A40 & A40 & GPU & Card & \cite{morgan2021nvidia}  \\ \hline  
     NVIDIA & Ampere A30 & A30 & GPU & Card & \cite{morgan2021nvidia}  \\ \hline  
     NVIDIA & Ampere A10 & A10 & GPU & Card & \cite{morgan2021nvidia}  \\ \hline  
     NVIDIA & Pascal P100 & P100 & GPU & Card & \cite{pascal2018nvidia,smith201816gb}  \\ \hline  
     NVIDIA & T4 & T4 & GPU & Card & \cite{kilgariff2018nvidia}  \\ \hline  
     NVIDIA & Volta V100 & V100 & GPU & Card & \cite{volta2019nvidia,smith201816gb}  \\ \hline  
     NVIDIA & DGX Station & DGX-Station & GPU & System & \cite{alcorn2017nvidia}  \\ \hline  
     NVIDIA & DGX-1 & DGX-1 & GPU & System & \cite{alcorn2017nvidia,cutress2018nvidias}  \\ \hline  
     NVIDIA & DGX-2 & DGX-2 & GPU & System & \cite{cutress2018nvidias}  \\ \hline  
     NVIDIA & DGX-A100 & DGX-A100 & GPU & System & \cite{campa2020defining}  \\ \hline  
     NVIDIA & Jetson TX1 & Jetson1 & GPU & System & \cite{franklin2017nvidia}  \\ \hline  
     NVIDIA & Jetson TX2 & Jetson2 & GPU & System & \cite{franklin2017nvidia}  \\ \hline  
     NVIDIA & Jetson Xavier NX & XavierNX & GPU & System & \cite{smith2019nvidia}  \\ \hline  
     NVIDIA & Jetson AGX Xavier & XavierAGX & GPU & System & \cite{smith2019nvidia}  \\ \hline  
     Perceive & Ergo & Perceive & dataflow & Chip & \cite{mcgregor2020perceive}  \\ \hline  
     PEZY Computing & PEZY-SC2 & PEZY-SC2 & manycore & System & \cite{schor2017pezy}  \\ \hline  
     Preferred Networks & MN-3 & Preferred-MN-3 & manycore & Card & \cite{preferred2020mncore, cutress2019preferred}  \\ \hline  
     Quadric & q1-64 & Quadric & dataflow & Chip & \cite{firu2019quadric}  \\ \hline  
     Qualcomm & Cloud AI 100 & Qcomm & dataflow & Card & \cite{ward2020qualcomm,mcgrath2019qualcomm}  \\ \hline  
     Rockchip & RK3399Pro & RK3399Pro & dataflow & Chip & \cite{rockchip2018rockchip}  \\ \hline  
     SiMa.ai & SiMa.ai & SiMa.ai & dataflow & Chip & \cite{gwennap2020machine}  \\ \hline  
     Syntiant & NDP101 & Syntiant & PIM & Chip & \cite{mcgrath2018tech,merritt2018syntiant}  \\ \hline  
     Tenstorrent & Tenstorrent & Tenstorrent & manycore & Card & \cite{gwennap2020tenstorrent}  \\ \hline  
     Tesla & Tesla Full Self-Driving Computer & Tesla & dataflow & System & \cite{talpes2020compute,wikichip2020fsd}  \\ \hline  
     Texas Instruments & TDA4VM & TexInst & dataflow & Chip & \cite{ward2020ti,ti2021tda4vm,demler2020ti}  \\ \hline  
     Toshiba & 2015 & Toshiba & multicore & System & \cite{merritt2019samsung}  \\ \hline  
     Untether & TsunAImi & TsunAImi & PIM & Card & \cite{gwennap2020untether} \\ \hline
     XMOS & xcore.ai & xcore.ai & dataflow & Chip & \cite{ward2020xmos}  \\ \hline

\end{tabular}
\end{table*}
\endgroup

Finally, a reasonable categorization of accelerators follows their intended application, and the five categories are shown as ellipses on the graph, which roughly correspond to performance and power consumption: Very Low Power for speech processing, very small sensors, etc.; Embedded for cameras, small UAVs and robots, etc.; Autonomous for driver assist services, autonomous driving, and autonomous robots; Data Center Chips and Cards; and Data Center Systems. 

We can make some general observations from Figure~\ref{fig:PeakPerformancePower}. First, a few new accelerator chips, cards, and systems have been announced and released in the past year. The density has clearly increased in the autonomous ellipse and data center cards and chips ellipse. Further, several cards and chips have been released that are focused on inference that exceed a peak power of 100W, e.g., Intel Habana Goya, NVIDIA Ampere A10 and A40, Alibaba, and Groq. This is a deviation from the last few years. This suggests that the power budget for autonomous vehicles and drones has crept past 100W, and that these accelerators are aimed at both the autonomous vehicle and data center inference markets. 
When it comes to precision, int8 has become the default numerical precision for embedded, autonomous and data center inference applications. Along with int8 for inference, a number of accelerators are also featuring fp16 and/or bf16 for inference. 
Finally, the competition for high-end training nodes shown in the data center systems ellipse is intensifying. NVIDIA and Cerebras have very high performing nodes, while Graphcore and Groq also have strong entries. Google TPUs and SambaNova also are competing in this space, but they have only been reporting multinode benchmark results, rather than single system peak capabilities.

\subsection{New Accelerators}

For most of the accelerators, their descriptions and commentaries have not changed since last year so please refer to last year's paper for descriptions and commentaries. There are, however, several new releases that were not covered by last year's paper that are covered here. In the following listings, the angle-bracketed string is the label of the item on the scatter plot, and the square bracket after the angle bracket is the literature reference from which the performance and power values came. 

\begin{itemize}
\item Blaize has emerged from stealth mode and announced its Graph Streaming Processor (GSP)~\cite{demler2020blaize}, but they have not provided any details beyond a high level component diagram of their chip. 
\item Enflame Technology, backed by Tencent, started shipping its CloudBlazer T10 data center training accelerator PCIe card~\cite{wheeler2021enflame,clarke2019globalfoundries}, which will support a broad range of datatypes including fp32, fp16, bf16, int32, int16, and int8. The T10 accelerator is focused on data center DNN training applications. 
\item Untether announced their TsunAImi card, which features four RunAI200 chips, during the Fall of 2020. Their at-memory design places 250,000 processing elements within a standard SRAM array. They are targeting the inference market and expect to ship cards in the first half of 2021. 
\item The Texas Instruments TDA4VM chip $\langle$TexInst$\rangle$ is a feature-rich automotive/autonomous system on a chip (SoC). It not only includes a 8 TOPS (int8) deep learning matrix multiply accelerator (MMA) with 4,096 computational units, but also eight ARM cores, two C7x vector DSPs, two C66x DSPs, 8 MB of L3 RAM, and several other audio, video, and security accelerators.~\cite{ward2020ti,ti2021tda4vm,demler2020ti} 
\item Mobileye, a subsidiary of Intel, has released its fifth generation automotive AI processor, EyeQ5 $\langle$EyeQ5$\rangle$. It includes eight CPU cores and 18 computer vision AI processors~\cite{eetimes2018mobileye,demler2020blaize}.  
\item The updated Mythic Intelligent Processing Unit accelerator $\langle$Mythic76$\rangle$~\cite{fick2018mythic,hemsoth2018mythic} combines a RISC-V control processor, router, and flash memory that uses variable resistors (i.e., analog circuitry) to compute matrix multiplies. The accelerators are aiming for embedded, vision, and data center applications. It is a smaller, lower-power 76 sq.mm. version of the 108 sq.mm., which is labeled $\langle$Mythic108$\rangle$.
\item Qualcomm has announced their Cloud AI 100 accelerator $\langle$QComm$\rangle$~\cite{mcgrath2019qualcomm}, and with their experience in developing communications and smartphone technologies, they have the potential for releasing a chip that delivers high performance with low power draws. 
\item Several new NVIDIA Ampere data center GPU cards have been released in the past year. The Ampere A40 $\langle$A40$\rangle$ and A10 $\langle$A10$\rangle$ are follow-on GPUs for data center inference to the Turing T4 card, while the Ampere A30 $\langle$A30$\rangle$ is a lower-performance, lower-power, more affordable version of the Ampere A100 training and HPC card~\cite{morgan2021nvidia}. 
\item In June 2021, Google shared details about their fourth generation inference-only  TPU4i accelerator $\langle$TPU4i$\rangle$~\cite{jouppi2021ten}. It features four 16k-element systolic matrix multiply units and was first deployed in early 2020. As with previous TPU variants, TPU4i is available through the Google Compute Cloud and for internal operations. 
\item Cerebras partnered with the TSMC chip fabrication foundry to scale its tiled wafer scale engine (WSE) from 16-nm feature size to 7-nm. The result is the Cerebras CS-2 $\langle$CS-2$\rangle$~\cite{trader2021cerebras}, which has 850,000 simple arithmetic units. On-board memory scales commensurately along with local memory bandwidth and internal bandwidth. The CS-2 chassis is the same 15-U rack mount system and 12 100-GigE network uplinks, and the system draws up to 23 KW of power. 

\end{itemize}

Finally, we must mention three accelerators that do not appear on Figure~\ref{fig:PeakPerformancePower} yet. Each has been released with some benchmark results but either no peak performance numbers or no peak power numbers. 

\begin{itemize}
 
\item Graphcore has announced its second generation accelerator chip, the CG200, which they are offering in their M2000 IPU Machine computer node. The M2000 incorporates four CG200 accelerators and Graphcore reports that the M2000 is capable of over a petaflop/s of performance~\cite{toon2020introducing,lunden2020graphcore}. While Graphcore has released some training results for the MLPerf benchmark~\cite{russell2021latest}, they have not disclosed any peak power or peak performance values. 
\item SambaNova has released some impressive benchmark results for their reconfigurable AI accelerator technology, but they still have not provided any details from which we can estimate peak performance or power consumption of their solutions~\cite{ward2020sambanova}. 
\item The Centaur Technology CNS processor~\cite{henry2020high, gwennap2019centaur} includes eight x86 cores along with an integrated AI accelerator realized as a 4,096 byte-wide SIMD unit. The Centaur AI coprocessor (CT-AIC) will delivers peak performance of 20 TOPS with INT8 precision at 2.5 GHz and can also operate at FP16 and INT16 precisions, though at lower performance. Centaur has not published any power numbers, though~\cite{gwennap2019centaur} predicts peak power to be less than 85 Watts. 
\end{itemize}

% Others to track: Alibaba TensorChip - only gives ops/watt for their two chips
% Untether AI - at-memory computations, but they don't 
% Esperanto ML chip with ~1100 risc-v cores

Much in the same way we no longer showed most FPGA-based solutions in last year's survey, we  are leaving out the research oriented chips that have not found their way to commercial production this year.
Research chips including Eyeriss~\cite{chen2018eyeriss,chen2017eyeriss,sze2017efficient}, EIE~\cite{han2016eie}, Tetris~\cite{gao2017tetris}, Tianjic~\cite{pei2019towards}, the DianNao family~\cite{chen2016diannao}, Adapteva~\cite{olofsson2016epiphany, olofsson2015kickstarting}, and NeuFlow~\cite{farabet2011neuflow} have been important in showing various computational performance, energy efficiency, and computational accuracy gains that could be achieved with specialized accelerator architectures and circuitry.

%Langroudi paper on neuromorphic processors mainly for inference~\cite{langroudi2019digital}
%A paper by Chen, et.al.~\cite{chen2020survey} surveys many accelerator architectures that have been and are being explored in research and commercial chips. 
%Wang paper with many references~\cite{wang2019deep}
%\cite{khan2020ai}

\section{Computational Efficiency Analysis}

% A few of the performance values are reported in frames per second (fps) with a given machine learning model. For translating fps values to performance values, Samuel Albanie's Matlab code and a web site of all of the major machine learning models with their operations per epoch/inference, parameter memory, feature memory, and input size~\cite{albanie2019convnet} were used. 

In recent years, a number of companies have been reporting actual performance numbers for their chips, cards, and systems. They have been doing so in the context of various benchmarks including MLPerf~\cite{reddi2019mlperf}. Most of the benchmark results have been for inference, where the metrics are images/items per second for throughput and latency to result. There have also been some training benchmark results, where the metric is time to train a particular DNN model~\cite{russell2021latest}. Since fielded defense and national security applications rely heavily on inference, we will focus on inference this year. Further, we will focus on images per second throughput over latency because in our experience, current defense and national security applications often prioritize throughput rate because the images/items from the sensor platform are collected in a consistent stream. Also DNN inference is more straightforward to characterize from a computational and data motion perspective. 

Fortunately, the DNN models that are specified in these benchmarks are well defined; that is, computing an inference output for any image (or other input item), the models are consistent and deterministic in the computations and data motion. The most accessible analysis of a wide set of DNN/CNN models is the online document maintained by Dr. Samuel Albanie~\cite{albanie2019convnet}. His table reports the number of fused-multiply-add (FMA) operations that dominate the inference (forward pass) computation. However, because it only reports the FMA operations, we must keep in mind that it is only an approximation of all of the computations and data motions involved. With FMA computation per single batch inference from Dr. Albanie's table, we can compute an approximation of the number of operations per second from the reported number of images processed per second. This is captured in Table~\ref{tab:utilization}. 

Table~\ref{tab:utilization} is sorted by images-per-second (IPS), and the four Google entries at the bottom. In~\cite{jouppi2021ten}, Google reports average computational efficiency results across the eight most utilized models at Google. Almost all of the accelerators are achieving over 20 percent computational efficiency; often it is challenging to achieve 10 percent computational efficiency on dense computational kernels with reasonably high arithmetic intensity~\cite{williams2009roofline}. Considering the highly parallel design of these ML accelerators with wide data transfer paths, it is reasonable to expect that further tuning and optimization should bring the utilization of those below 20 percent higher using techniques including strategic data layouts and data transfer latency hiding. Interestingly, there is no correlation between technology type, precision, or application category with computational utilization percentage.

\begingroup
\setlength{\tabcolsep}{2pt} % Default value: 6pt

\begin{table*}
  \centering
  \caption{Inference utilization percentage for select accelerators.}
  \label{tab:utilization}

\begin{tabular}{| l | l | c | c | c | c | c | c | c | c |} \hline  \hline

 \textbf{Company/Org} & \textbf{Accelerator} & \textbf{Tech.} & \textbf{DNN Model} & \textbf{IPS} & \textbf{Perf. (TOPS)} & \textbf{Precision} & \textbf{Utilization Percent} & \textbf{References} \\ \hline 
     GreenWaves & GAP9 & dataflow & MobileNetV1 & 83.9 & 0.0489 & int8 & 46\% & \cite{turley2020gap9}  \\ \hline  
     % Google & TPUedge & dataflow & MobileNetV2 & 100 & 0.0585 & int8.32 & 1.5\% & \cite{tpu2019edge}  \\ \hline  
     NovuMind & NovuTensor & dataflow & ResNet-34 & 697 & 2.8 & int8 & 19\% & \cite{yoshida2018novumind}  \\ \hline  
     Mythic & Mythic108 & PIM & ResNet-50 & 900 & 3.6 & analog & 10\% & \cite{fick2018mythic}  \\ \hline  
     Nvidia & Jetson Xavier NX & GPU &  ResNext-50 & 1,165 & 4.7 & int8 & 22\% & MLPerf 1.0-98 \\ \hline 
     Nvidia & Jetson AGX Xavier & GPU &  ResNext-50 & 2,072 & 8.3 & int8 & 26\% & MLPerf 1.0-92 \\ \hline 
     Intel & 2xXeon8280 & manycore & ResNext-50 & 3,248 & 13.0 & int8 & 34\% & \cite{degelas2019intel}  \\ \hline  
     Nvidia & T4 & GPU & ResNet-50 & 4,292 & 17.2 & int8 & 13\% & \cite{harmon2019nvidia}  \\ \hline  
     Kalray & Coolidge & manycore & GoogleNet & 6,000 & 12 & int8.32 & 50\% & \cite{dupont2019kalray}  \\ \hline  
     Qualcomm & Cloud AI 100 & dataflow &  ResNext-50 & 7,807 & 31 & int8 & 8\% & MLPerf 1.0-101 \\ \hline 
     Nvidia & V100 & GPU & ResNet-50 & 7,907 & 31.6 & int8 & 56\% & \cite{gwennap2020groq}  \\ \hline  
     Achronix & Achronix & dataflow & ResNet-50 & 8,600 & 34.4 & int8 & 40\% & \cite{roos2019fpga}  \\ \hline  
     Cambricon & MLU100 & dataflow & ResNet-50 & 10,000 & 40 & int8 & 31\% & \cite{peng2019alibaba}  \\ \hline  
     Habana & Goya & dataflow & ResNet-50 & 15,433 & 61.7 & int8 & 62\% & \cite{gwennap2019habanagoya}  \\ \hline  
     Groq & TSP & dataflow & ResNet-50 & 21,700 & 86.8 & int8 & 11\% & \cite{krewell2020virtual,gwennap2020groq}  \\ \hline  
     Tenstorrent & Tenstorrent & manycore & ResNet-50 & 22,431 & 89.7 & int8 & 24\% & \cite{gwennap2020tenstorrent}  \\ \hline  

     Nvidia & A100 & GPU &  ResNext-50 & 38,010 & 152 & int8 & 24\% & MLPerf 1.0-29 \\ \hline

     Google & TPU1 & dataflow & Google-8 & -- & -- & int8 & 20\% & \cite{jouppi2021ten} \\ \hline
     Google & TPU2 & dataflow & Google-8 & -- & -- & fp16 & 51\% & \cite{jouppi2021ten} \\ \hline
     Google & TPU3 & dataflow & Google-8 & -- & -- & bf16 & 38\% & \cite{jouppi2021ten} \\ \hline
     Google & TPU4i & dataflow & Google-8 & -- & -- & bf16 & 33\% & \cite{jouppi2021ten} \\ \hline

\end{tabular}
\end{table*}
\endgroup

\section{Summary}

This paper updated the survey of deep neural network accelerators that span from extremely low power through embedded and autonomous applications to data center class accelerators for inference and training. We focused on inference accelerators, and discussed some new additions for the year. The rate of announcements and releases has slowed down some, but we are starting to see second generation accelerators that are significantly improving on the capabilities of the first generation. Actual performance benchmark results are being released more, which gives us the opportunity to evaluate computational efficiency for the first time for accelerators for which we have benchmark results and peak performance numbers. Many of these accelerators achieve over 20 percent computational efficiency. 

\section{Data Availability}

The data spreadsheets and references that have been collected for this study and its papers will be posted at \url{https://github.com/areuther/ai-accelerators} after they have cleared the release review process. 

\section*{Acknowledgement}

We are thankful to Masahiro Arakawa, Bill Arcand, Bill Bergeron, David Bestor, Bob Bond, Chansup Byun, Nathan Frey, Vitaliy Gleyzer, Jeff Gottschalk, Michael Houle, Matthew Hubbell, Hayden Jananthan, Anna Klein, David Martinez, Joseph McDonald, Lauren Milechin, Sanjeev Mohindra, Paul Monticciolo, Julie Mullen, Andrew Prout, Stephan Rejto, Antonio Rosa, Matthew Weiss, Charles Yee, and Marc Zissman
for their support of this work. 


\bibliographystyle{IEEEtran} 
\bibliography{MLAcceleratorAnalysis}

%%\end{thebibliography}

\end{document}